\documentclass[aps,prb,reprint,amsmath,amssymb,superscriptaddress,floatfix,longbibliography]{revtex4-2}
\usepackage{graphicx}
\usepackage{bm}
\usepackage{natbib}
\usepackage{braket}
\usepackage{mathtools}
\usepackage{xcolor}
\usepackage[colorlinks,bookmarks=true,citecolor=blue,linkcolor=blue,urlcolor=blue]{hyperref}
\IfFileExists{orcidlink.sty}{\usepackage{orcidlink}}{}
\providecommand{\orcidlink}[1]{\href{https://orcid.org/#1}{\textsuperscript{\textsf{iD}}}}

\newcommand{\ii}{\mathrm{i}}

\begin{document}

\title{Reconstructing chiral edge states via boundary dissipation}

\author{Xin-Ran Ma\orcidlink{0009-0007-4051-7417}}
\affiliation{Center for Advanced Quantum Studies, School of Physics and Astronomy, Beijing Normal University, Beijing 100875, China}

\author{Kui Cao}
\affiliation{Center for Advanced Quantum Studies, School of Physics and Astronomy, Beijing Normal University, Beijing 100875, China}

\author{Qian Du}
\affiliation{School of Physics and Electronic Engineering, Linyi University, Linyi 276000, China}

\author{Zheng Wei}
\affiliation{Center for Advanced Quantum Studies, School of Physics and Astronomy, Beijing Normal University, Beijing 100875, China}

\author{Su-Peng Kou}
\email{spkou@bnu.edu.cn}
\affiliation{Center for Advanced Quantum Studies, School of Physics and Astronomy, Beijing Normal University, Beijing 100875, China}

\date{\today}

\begin{abstract}
Boundary dissipation can reshape the spatial profiles and decay properties of topological edge modes in Chern insulators. We develop a microscopic boundary-channel framework to explain their decay mechanisms and determine the effective decay rates in the weak- and strong-loss regimes. By retaining the lossy outer orbitals and a normalized edge state of the Hermitian inner system, we identify a directly coupled bright boundary channel, while incorporating the remaining inner states through a Feshbach self-energy. In the Qi--Wu--Zhang model, weak-loss decay is governed to leading order by the original edge state's probability weight on the lossy boundary, with coupling-induced corrections starting at cubic order. At strong loss, the reconstructed inner-edge state instead decays through virtual coupling to the damped bright channel, giving a rate proportional to the squared microscopic coupling and inversely proportional to the bare loss. The Haldane model shows that the same leading decay mechanism holds for a different intercell hopping structure, while additional matrix-valued self-energy feedback modifies the boundary spectrum at finite loss and shifts the exceptional-point threshold. These results provide a unified microscopic description of boundary-state reconstruction and its decay mechanisms, offering testable predictions to guide experiments in photonic crystals, phononic crystals, and topolectrical circuits.
\end{abstract}

\maketitle
\raggedbottom

\section{Introduction}
\label{sec:introduction}
Chiral edge states of Chern insulators are protected against elastic backscattering by bulk topology~\cite{RevModPhys.82.3045,R-TKNN,R-Hatsugai}, making them promising for robust transport. Yet topological protection neither fixes their spatial profiles nor prevents dissipative decay, making control of these properties important for tailoring edge-state coupling and transport. Dissipation can thus serve as a means of control rather than merely a source of decay. In open quantum systems, dissipative dynamics are commonly described by the Lindblad master equation~\cite{Breuer_oqs,PhysRevB.102.115109,RevModPhys.94.045006}; within the quantum-trajectory framework, evolution conditioned on no quantum jumps is governed by an effective non-Hermitian Hamiltonian. Combined with nontrivial band topology, dissipation can stabilize or select interface modes and prepare topological states~\cite{Poli2015,Weimann2017,PhysRevLett.131.083801,Slootman2024Monomode,Verstraete2009,Diehl2011,Bardyn2013,Budich2015,PhysRevResearch.5.043229,Peng2026}. It can also reshape topological mode profiles and transport~\cite{Zhao_2019,R-Tang2025,Wang2022Morphing,Fan2023Reconfigurable,Xie2025ValleyLifetime,Yin2026Valley}, while modifying edge damping, localization, and bulk--boundary detachment~\cite{PhysRevLett.131.256601,Sarkar2026,PhysRevLett.132.113802,PhysRevLett.135.266602}. Spatially patterned dissipation further extends this control to boundary and corner responses in two-dimensional lattices~\cite{PhysRevLett.123.073601,PhysRevLett.128.223903,PhysRevB.106.035425,Zou2021,Jiang2024BoundaryHSTE,Wu2024Helical}.

Boundary dissipation provides a direct setting for studying the interplay between topology and non-Hermiticity in Chern insulators, as it introduces loss on the outermost orbitals while leaving the Hermitian bulk Hamiltonian unchanged~\cite{RevModPhys.94.045006,PhysRevLett.131.256602,PhysRevResearch.6.013213}. Previous studies have explored the non-Hermitian skin effect of chiral edge states~\cite{PhysRevResearch.6.013213} and non-Hermitian boundary topology~\cite{PRXQuantum.4.030315,PhysRevLett.131.256602}. Non-Hermitian spectra can also exhibit exceptional points (EPs), where eigenvalues and eigenvectors coalesce~\cite{RevModPhys.93.015005,Heiss_2012,Dembowski2001}. Studies of non-Hermitian domain walls, dissipation-induced state separation, and proximity effects further highlight how non-Hermiticity can reshape the spatial profiles of states~\cite{PhysRevLett.129.053903,Du2024,Wu2023,He2026CSE}.

For a Chern insulator with uniform dissipation on its outermost boundary orbitals, chiral edge states exhibit a counterintuitive behavior: their decay rate increases linearly with loss strength in the weak-loss regime~\cite{PhysRevResearch.6.013213}, but decreases inversely in the strong-loss regime~\cite{Lu2024}. How can stronger dissipation suppress the decay of an edge state that originally resides near the lossy boundary? Although the asymptotic decay-rate scalings are known in the weak- and strong-loss limits, a unified microscopic framework connecting these two behaviors through edge-state reconstruction and coupling to the dissipative boundary remains lacking.

In this work, we develop a microscopic boundary-channel framework that retains the lossy outer orbitals and a normalized edge state of the Hermitian inner system.
Hopping between them selects one directly coupled outer combination, the bright state, while the remaining inner states enter through a Feshbach self-energy~\cite{Feshbach1958,Feshbach1962,Rotter2009}.
For the Qi--Wu--Zhang (QWZ) model, an exact unitary transformation connects this basis to one containing the original Hermitian edge state and reveals distinct decay mechanisms in the two loss regimes.
For an isolated QWZ edge mode, weak boundary loss of strength $\gamma$ gives a leading decay rate $\Gamma_{\mathrm{weak}}\simeq\gamma Z_k$, where $Z_k$ is the original edge state's probability weight on the lossy boundary; coupling-induced corrections to the decay rate are of order $\gamma^3$.
At strong loss, the reconstructed inner edge decays through virtual occupation of the damped bright state, giving $\Gamma_{\mathrm{strong}}\simeq G_k^2/\gamma$, where $G_k$ is fixed by the normalized edge wave function and microscopic hopping.
At resonance, the bright and inner states yield an exactly solvable EP.
For the Haldane model, rank-two intercell hopping allows matrix-valued self-energy feedback from the remaining inner states, which shifts the EP. Our framework provides a quantitative microscopic description of chiral edge-state decay and localization.

\section{Dissipation-induced Boundary reconstruction}
\label{sec:phenomenon}
\begin{figure}[tbp]
\centering
\includegraphics[width=\columnwidth]{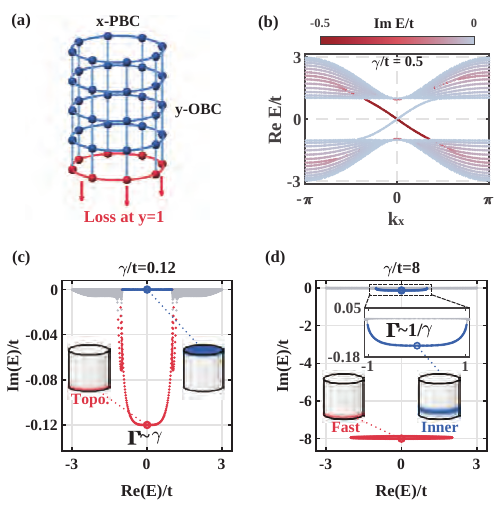}
\caption{Boundary reconstruction in the QWZ cylinder. (a) Schematic of boundary loss applied to both orbitals at $y=1$. (b) Dispersion of $\operatorname{Re}E$ at $\gamma/t=0.5$, with color indicating $\operatorname{Im}E/t$. (c),(d) Complex-energy spectra at $\gamma/t=0.12$ and $8$, respectively. Red denotes the lossy outer-edge branches, blue denotes the opposite-edge branch and, in (d), the reconstructed inner-edge branch, and gray denotes the remaining states. The sketches illustrate the spatial profiles of the modes at $k_x=0$. The inset in (d) magnifies the reconstructed branch, whose decay rate follows $\Gamma\sim t^2/\gamma$ in the strong-loss limit. The parameters are $M/t=-1$ and $N_y=40$.}
\label{fig:reconstruction}
\end{figure}

We consider the QWZ model on a cylinder~\cite{QWZ2006}, periodic along $x$ and open along $y$ [Fig.~\ref{fig:reconstruction}(a)]. Equal loss acts on both orbitals of the outer row $y=1$:
\begin{align}
H(k)={}&
\sum_{y=1}^{N_y}
|y\rangle\langle y|\otimes h(k)
\nonumber\\
&+\sum_{y=1}^{N_y-1}
\Bigl(
|y\rangle\langle y+1|\otimes T
+\mathrm{H.c.}
\Bigr)
-\ii\gamma P_1,
\label{eq:qwzstrip}
\end{align}
where $k\equiv k_x$ is the momentum along $x$, $t$ is the hopping amplitude, $M$ is the orbital mass, $\gamma\geq0$ is the boundary loss strength, and we define the shorthand notations $a_k=t\sin k$, $b_k=M+t\cos k$, $P_1=|1\rangle \langle 1| \otimes\sigma_0$, $h(k)=a_k\sigma_x+b_k\sigma_z$, $T=0.5t(\sigma_z-\ii\sigma_y)$. We take $M = -t$ and set $\hbar$ and the lattice spacing to unity.

At $\gamma/t=0.5$, Fig.~\ref{fig:reconstruction}(b) shows two chiral edge branches connecting the lower and upper bulk bands in the gap. They are localized at opposite edges. The color reveals their different imaginary energies: the branch on the lossy edge has $\operatorname{Im}E<0$, while the opposite-edge branch remains nearly real. Figure~\ref{fig:reconstruction}(c) resolves this separation in the complex spectrum at weaker loss, $\gamma/t=0.12$.

At $\gamma/t=8$, rapidly decaying outer branches with $\operatorname{Im}E\sim-\gamma$ coexist with a slowly decaying chiral inner-edge branch close to the real axis [Fig.~\ref{fig:reconstruction}(d)]. The spatial sketches at $k_x=0$ locate the fast modes on $y=1$ and the slow mode near $y=2$. Strong loss thus effectively excludes the outer row from the slow dynamics, exposing a topological boundary one row inward. We call this inward shift \emph{\textit{boundary reconstruction}}.

The negative imaginary energy gives the decay rate of a right eigenstate $\ket{\Phi_{\mu}}$:
\begin{equation}
\Gamma_\mu \equiv-\operatorname{Im}E_\mu=\gamma\frac{\bra{\Phi_\mu}P_1\ket{\Phi_\mu}}{\braket{\Phi_\mu|\Phi_\mu}}.
\label{eq:rate}
\end{equation}
The ratio is the state's probability on the lossy row, so $\Gamma_\mu$ measures its effective dissipation. The original edge has $\Gamma \propto \gamma$ at weak loss, whereas the reconstructed branch has $\Gamma \propto \gamma^{-1}$ at strong loss. The inset of Fig.~\ref{fig:reconstruction}(d) resolves its small but finite decay rate; its lifetime increases with the boundary loss strength.

\section{Microscopic theory of chiral edge reconstruction}
\label{sec:theory}
Strong boundary loss separates rapidly damped outer orbitals from a slowly decaying edge mode of the remaining inner system[Fig.~\ref{fig:reconstruction}(d)]. To construct a basis for the reconstructed boundary, we first remove the lossy outer layer from the Hermitian Hamiltonian $H_0$ at each conserved momentum and denote the remaining Hamiltonian by $H_{\rm in}$. We solve $H_{\rm in}\ket{t_Q}=\epsilon_Q\ket{t_Q}$ for an isolated, normalized inner-edge state and retain it together with all $N_s$ outer orbitals ${\ket{d_\alpha}}$. The projectors are
\begin{equation}
P=P_1+\ket{t_Q}\bra{t_Q},\qquad Q=I-P.
\label{eq:projectors}
\end{equation}
The reference state $\ket{t_Q}$ belongs to the retained $P$ sector; the projector $Q$ eliminates only the other inner states. The orbital spinors $u_y$ of $\ket{t_Q}$ on layers $y\geq2$ obey the global normalization $\sum_{y\geq2}u_y^\dagger u_y=1$. The first spinor $u_2$ alone generally has norm smaller than one.

We call the outer combination that couples directly to this inner edge the bright state $|B\rangle$ and construct it from the microscopic coupling:
\begin{equation}
\begin{gathered}
g_\alpha=\bra{d_\alpha}H_0\ket{t_Q},\qquad
G=(g^\dagger g)^{1/2},\\
\ket B=\frac{1}{G}\sum_{\alpha=1}^{N_s}g_\alpha\ket{d_\alpha}.
\end{gathered}
\label{eq:bright}
\end{equation}
For nearest-layer hopping, $g=T_{\rm out,in}u_2$, where $T_{\rm out,in}=\langle1|H_0|2\rangle$ maps the first inner layer to the outer layer. We therefore obtain $G$ from the hopping and the globally normalized edge wave function. For $G>0$, we complete $B$ to an orthonormal outer basis by adding $N_s-1$ dark states $D_j$ satisfying $\langle D_j|B\rangle=0$. Then $\langle B|H_0|t_Q\rangle=G$ and $\langle D_j|H_0|t_Q\rangle=0$. A single retained edge state selects one bright state, regardless of the rank of the interlayer hopping. These dark states have no direct coupling to the retained inner edge, but they still experience the loss $-\ii\gamma$.

\begin{figure}[tbp]
\centering
\includegraphics[width=\columnwidth]{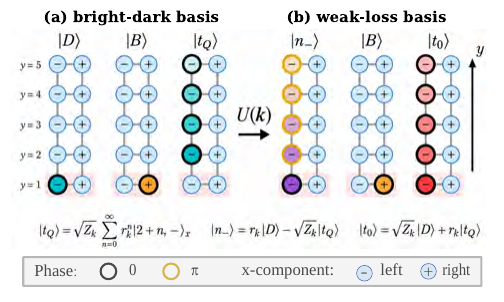}
\caption{Boundary bases in the QWZ model. (a) Bright-dark (strong-loss) basis $(D,B,t_Q)$, where $D$ and $B$ are lossy outer-row states and only $B$ couples directly to the inner-edge state $t_Q$. (b) Weak-loss basis $(n_-,B,t_0)$, obtained through the unitary transformation $U(k)$, with $t_0$ denoting the original topological edge state. Filled areas indicate orbital weights, outline colors indicate phases, and red shading marks the lossy row.}
\label{fig:basis}
\end{figure}

Figure~\ref{fig:basis}(a) shows this basis for the QWZ model. Its two outer orbitals form a bright $|B\rangle$ and a dark state$|D\rangle$, while the inner reference state starts at the next layer. The inner state decays by coupling to the lossy bright state. Boundary mixing can also connect $B$ to $D$. To calculate the outer-state energies and mixing, we project the Hermitian outer Hamiltonian $h_{\rm out}=P_1H_0P_1$:
\begin{equation}
h_{mn}=\bra mh_{\rm out}\ket n,\qquad
\widetilde h_{mn}(E)=h_{mn}+\Sigma^{\rm reg}_{mn}(E),
\label{eq:hdefinitions}
\end{equation}
where $m,n\in\{B,D_1,\ldots,D_{N_s-1}\}$. The residual self-energy is the correction to the retained Hamiltonian produced by the remaining inner states. We obtain this Feshbach self-energy by eliminating those inner states:
\begin{equation}
\Sigma^{\rm reg}(E)=PH_0Q(E-QH_0Q)^{-1}QH_0P.
\label{eq:feshbach}
\end{equation}
Because $\ket{t_Q}$ is an eigenstate of $H_{\rm in}$, $QH_0\ket{t_Q}=0$; this self-energy acts only within the outer block. We evaluate this self-energy from the Hermitian inner resolvent, the inverse of energy minus the inner Hamiltonian, or by summing over the other inner eigenstates, as derived in Appendix~\ref{app:general}. The self-energy depends on $\gamma$ only through the eigenvalue $E$ at which we evaluate it.

In the basis order $(B,D_1,\ldots,D_{N_s-1},t_Q)$, we obtain
\begin{equation}
H_{\rm bd}(E)=
\begin{pmatrix}
\widetilde h_{BB}-\ii\gamma&\widetilde h_{BD}&G\\
\widetilde h_{DB}&\widetilde h_{DD}-\ii\gamma I_D&0\\
G&0&\epsilon_Q
\end{pmatrix}.
\label{eq:generalbd}
\end{equation}
Here $D$ denotes all dark states, $I_D$ is the identity in that sector, and entries connecting it to the other states are row or column vectors. We find the boundary modes by solving $H_{\rm bd}(E)u=Eu$ with the energy-dependent self-energy $\Sigma^{\rm reg}(E)$. This equation is exact when the eliminated resolvent exists and the eigenstate has nonzero weight in the retained space. An energy-independent few-state model additionally requires a controlled approximation to the self-energy.

To find the inner-edge pole, we eliminate the outer amplitudes:
\begin{equation}
E-\epsilon_Q
=G^2\left[(E-\widetilde h(E)+\ii\gamma I)^{-1}\right]_{BB}.
\label{eq:innerpole}
\end{equation}
At sufficiently strong loss and for an isolated inner-edge branch, the bright amplitude is approximately $-\ii G/\gamma$ times the inner amplitude. Consequently its outer weight is $G^2/\gamma^2$ to leading order, and Eq.~\eqref{eq:rate} gives $\Gamma_{\rm in}\simeq G^2/\gamma$.  Increasing the bare loss suppresses this transfer and lengthens the lifetime of the reconstructed mode.

To expose the weak-loss mechanism, we rotate the Hamiltonian within the same retained space, $H'=U^\dagger H_{\rm bd}U$, transforming both the loss and the self-energy. Figure~\ref{fig:basis}(b) shows the resulting QWZ basis containing the original topological edge state; Sec.~\ref{sec:qwz} analyzes its weak-loss decay. Detailed proofs are shown in Appendix~\ref{app:general}.

\section{Distinct weak- and strong-loss decay mechanisms in the QWZ model}
\label{sec:qwz}
For the QWZ model, we define $r_k=-b_k/t$, and $Z_k=1-r_k^2$. For $|r_k|<1$, the Hermitian
inner system has the normalized edge state
\begin{equation}
\begin{gathered}
\ket{t_Q(k)}=\sqrt{Z_k}\sum_{n=0}^{\infty}r_k^n\ket{2+n,-}_x,
\qquad \epsilon_Q=-a_k,\\
\ket B=\ket{1,+}_x,\qquad \ket D=\ket{1,-}_x,
\qquad G_k=t\sqrt{Z_k}.
\end{gathered}
\label{eq:qwzcoupling}
\end{equation}
Here $\pm$ denote the eigenstates of $\sigma_x$. $Z_k$ is the probability weight of $|t_Q\rangle$ on $y=2$; the rank-one transverse hopping couples this component to the bright outer state $|B\rangle$. The orthogonal outer state $|D\rangle$ does not couple directly to $|t_Q\rangle$.
In the strong-loss basis $(B,D,t_Q)$, the effective model is
\begin{equation}
H_{\rm bd}^{\rm Q}(E)=
\begin{pmatrix}
a_k-\ii\gamma+\Sigma^{\rm reg}_{BB}&b_k&G_k\\
b_k&-a_k-\ii\gamma&0\\
G_k&0&-a_k
\end{pmatrix}.
\label{eq:qwzbd}
\end{equation}
The residual self-energy accounts for coupling to Hermitian inner states other than $|t_Q\rangle$. It acts only on $|B\rangle$ because the interlayer hopping has rank one.
Equation~\eqref{eq:qwzbd} remains exact beyond the strong-loss limit when we retain the energy dependence of the self-energy. Appendix~\ref{app:qwz} derives the self-energy and specifies the physical branch of the surface resolvent.

\begin{figure*}[!t]
\centering
\includegraphics[width=\textwidth]{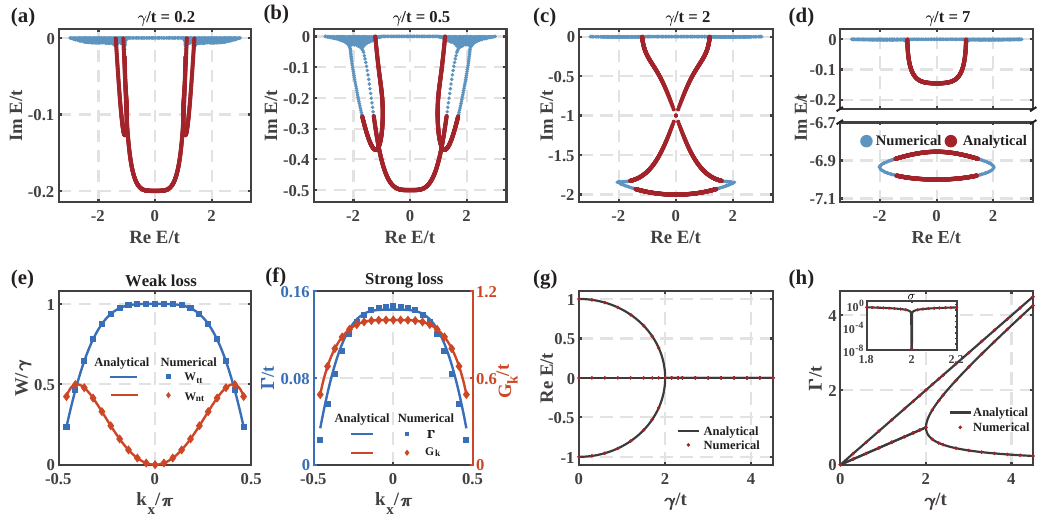}
\caption{Energy spectra and decay mechanisms in the QWZ model. (a)--(d) Energy spectra (blue) and analytical edge-state spectra (red) at $\gamma/t=0.2,0.5,2,7$, respectively. (e) Weak-loss direct-decay coefficient $W_{tt}/\gamma=Z_k$ and dissipative coupling $|W_{nt}|/\gamma=|r_k|\sqrt{Z_k}$. (f) Strong-loss decay rate $\Gamma/t\simeq tZ_k/\gamma$ and bright--inner coupling strength $G_k/t=\sqrt{Z_k}$ at $\gamma/t=7$. (g),(h) Real energies and decay rates of the three $k_x=0$ channels. The inset shows the smallest singular value $\sigma=s_{\rm min}(V)$, which vanishes at the EP. The parameters are $M/t=-1$, $N_y=40$, with $x$-PBC and $y$-OBC.}
\label{fig:qwzresults}
\end{figure*}

 The inner-edge state has no bare loss, but its coupling $G_k$ gives it a bright-state amplitude of order $G_k/\gamma$. Loss on the outer component produces
\begin{equation}
\Gamma_{\rm strong}(k)=\frac{G_k^2}{\gamma}
+O(\gamma^{-3})=\frac{t^2Z_k}{\gamma}+O(\gamma^{-3}).
\label{eq:qwzstrongrate}
\end{equation}
This is decay through virtual occupation of a strongly damped channel.
Increasing the bare loss suppresses this occupation and lengthens the lifetime of the inner-edge state. Figure~\ref{fig:qwzresults}(f) compares $G_k/t=\sqrt{Z_k}$ (orange) with $\Gamma/t\simeq tZ_k/\gamma$ (blue) at $\gamma/t=7$. The coupling $G_k$ sets the leading strong-loss decay rate through $G_k^2/\gamma$.

To identify the weak-loss mechanism, we express the same retained space in the order $(D,B,t_Q)$ of Fig.~\ref{fig:basis} and apply the unitary transformation
\begin{gather}
U=\begin{pmatrix}
r_k&0&\sqrt{Z_k}\\
0&1&0\\
-\sqrt{Z_k}&0&r_k
\end{pmatrix},\qquad U^\dagger U=I,
\label{eq:U}\\
\begin{aligned}
\ket{n_-}&=r_k\ket D-\sqrt{Z_k}\ket{t_Q},\\
\ket{t_0}&=\sqrt{Z_k}\ket D+r_k\ket{t_Q},
\end{aligned}
\label{eq:weakstates}
\end{gather}
where $|t_0\rangle$ is the original Hermitian edge state of the complete lattice, with probability $Z_k$ on $y=1$; $|n_-\rangle$ is its orthogonal companion. In the order $(n_-,B,t_0)$, the transformed model is
\begin{equation}
H_{\rm w}(E)=
\begin{pmatrix}
-a_k-\ii\gamma r_k^2&-t&-\ii\gamma r_k\sqrt{Z_k}\\
-t&a_k-\ii\gamma+\Sigma^{\rm reg}_{BB}&0\\
-\ii\gamma r_k\sqrt{Z_k}&0&-a_k-\ii\gamma Z_k
\end{pmatrix}.
\label{eq:qwzweak}
\end{equation}
The weak-loss approximation enters when we expand the eigenvalues of $H_{\rm w}$ at small $\gamma$. The transformed loss matrix $W=\gamma U^\dagger\operatorname{diag}(1,1,0)U$ has the entries
\begin{equation}
W_{tt}=\gamma Z_k,\qquad W_{nt}=\gamma r_k\sqrt{Z_k}.
\label{eq:Wentries}
\end{equation}
Thus $|t_0\rangle$ has direct diagonal loss. Its coupling to
$n_-$ is dissipative, with matrix element $-\ii W_{nt}$. For an isolated edge pole at fixed $k$, eliminating $(n_-,B)$ shows
that this coupling first shifts the real energy at order $\gamma^2$. And its first contribution to the decay rate is of order $\gamma^3$.
Explicitly,
\begin{equation}
\Gamma_{\rm weak}(k)=\gamma Z_k
+\frac{4a_k^2r_k^2}{t^4}\gamma^3+O(\gamma^5).
\label{eq:qwzweakrate}
\end{equation}
The boundary weight of the original topological edge state therefore sets the leading weak-loss decay rate. Coupling to the companion states gives a higher-order correction, whereas coupling to $|B\rangle$ generates the leading decay of the strong-loss inner-edge state.

Figure~\ref{fig:qwzresults}(e) and (f) distinguish these
mechanisms through their microscopic matrix elements. Figure~\ref{fig:qwzresults}(e) compares the direct-loss coefficient $W_{tt}/\gamma=Z_k$ with the dissipative mixing $|W_{nt}|/\gamma=|r_k|\sqrt{Z_k}$; although the mixing can be larger at some momenta, it contributes to the decay rate only at higher order in the weak-loss expansion. Figure~\ref{fig:qwzresults}(f) instead relates the strong-loss decay to a
coherent, loss-independent coupling $G_k$. The common factor $Z_k$ in the two leading rates does not imply the same decay mechanism.

In Figs.~\ref{fig:qwzresults}(a)--(d), we obtain the energies of edge states from the effective model with the physical half-space poles of Eq.~\eqref{eq:Bsecular} and the energy-dependent $\Sigma^{\mathrm{Reg}}$. At intermediate loss or near an edge-continuum merger, we solve the full pole equation. Equations~\eqref{eq:qwzstrongrate} and \eqref{eq:qwzweakrate} apply in their respective limits.

An exactly solvable EP occurs at $M=-t$ and $k=0$. Here
$a_k=b_k=r_k=0$, $Z_k=1$, and $\Sigma^{\mathrm{Reg}}=0$,
so the retained model becomes
\begin{equation}
H_{\rm EP}=(-\ii\gamma)_D\oplus
\begin{pmatrix}-\ii\gamma&t\\t&0\end{pmatrix}_{B,t_Q}.
\label{eq:qwzblock}
\end{equation}
Thus $E_\pm=-\ii\gamma/2\pm\sqrt{t^2-\gamma^2/4}$, $\gamma_{\rm EP}=2t$, and $E_{\rm EP}=-\ii t$. The bright--inner eigenvectors coalesce at the EP. Above the EP, the pair separates into a fast branch with decay rate of order $\gamma$ and a slow branch whose rate approaches $t^2/\gamma$ [Figs.~\ref{fig:qwzresults}(g),(h)]. Meanwhile, the original edge state $|D\rangle$ ($|t_0\rangle$) remains an exact eigenstate with $E=-\ii\gamma$ and does not participate in the EP.

\begin{figure}[tbp]
\centering
\includegraphics[width=\columnwidth]{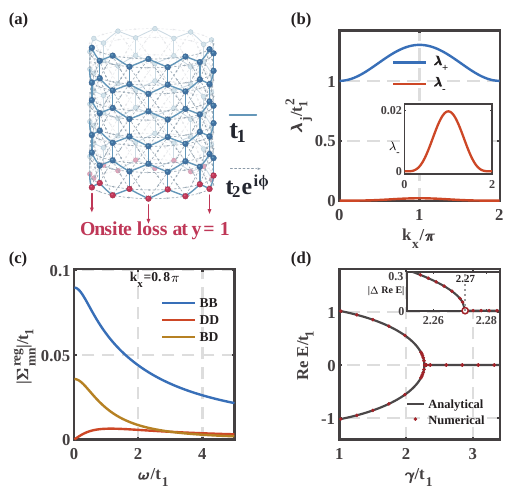}
\caption{Boundary channels and spectra in the Haldane model. (a) Schematic of boundary dissipation in the Haldane model with zigzag edges, showing nearest-neighbor hopping $t_1$, next-nearest-neighbor hopping $t_2e^{\ii\phi}$, and onsite loss at $y=1$ (red). (b) Squared intercell coupling strengths $\lambda_\pm/t_1^2$; the inset highlights the weaker coupling channel $\lambda_-$. 
(c) Residual self-energy elements $|\Sigma^{\rm reg}_{mn}|/t_1$ ($m,n\in\{B,D\}$) at $k_x=0.8\pi$, evaluated along $E=\epsilon_Q-\ii\omega$. (d) Real parts of the bright--inner boundary poles at $k_x=\pi$. The inset shows the splitting near the EP at $\gamma_{\rm EP}/t_1=2.27$. The parameters: $t_2/t_1=0.2$, $M=0$, $\phi=\pi/2$, and $N_y=40$.}
\label{fig:haldane}
\end{figure}

\section{Self-energy effects in the Haldane model}
\label{sec:haldane}
Boundary reconstruction in the Haldane model also produces an inner-edge mode whose decay comes from virtual coupling to the lossy boundary. Unlike QWZ, the zigzag Haldane edge has a generically rank-two intercell hopping matrix, allowing the remaining inner states to contribute to all entries of the boundary self-energy. The Haldane Hamiltonian~\cite{Haldane1988} with boundary loss reads
\begin{equation}
\begin{aligned}
H_{\rm Haldane}(k)={}&
\sum_{y=1}^{N_y}|y\rangle\langle y|\otimes h(k)\\
&+\sum_{y=1}^{N_y-1}
\left[|y\rangle\langle y+1|\otimes T(k)
+\mathrm{H.c.}\right]\\
&-\ii\gamma|1\rangle\langle1|\otimes I_2,
\end{aligned}
\label{eq:haldane_boundary_H}
\end{equation}
where $k$ is the momentum along the periodic direction, and $y$ labels zigzag cells from the lossy boundary inward, and the last term imposes uniform loss $-\ii\gamma$ on both sublattice orbitals $(A_1,B_1)$ on the zigzag boundary ($y=1$). We set nearest-neighbor hopping $t_1$, next-nearest-neighbor hopping $t_2e^{\ii\phi}$ with $t_2/t_1=0.2$, $M=0$, $\phi=\pi/2$[Fig.~\ref{fig:haldane}(a)]. In the $(A_y,B_y)$ sublattice basis,
\begin{equation}
\begin{gathered}
h(k)=
\begin{pmatrix}
2t_2\sin k & t_1(1+e^{-\ii k})\\
t_1(1+e^{\ii k}) & -2t_2\sin k
\end{pmatrix},\\[3pt]
T(k)=
\begin{pmatrix}
\ii t_2(1-e^{-\ii k}) & 0\\
t_1 & -\ii t_2(1-e^{-\ii k})
\end{pmatrix}.
\end{gathered}
\label{eq:haldane_boundary_blocks}
\end{equation}
Here $T(k)=\langle y|H_0|y+1\rangle$ is the upper
intercell hopping block, and the loss acts equally on
both $A_1$ and $B_1$.

To describe the reconstructed edge, we remove the first cell and define $u_Q=(u_A,u_B)^T$ as the amplitude of the normalized inner-edge state $t_Q$ on the newly exposed cell. The microscopic couplings are
\begin{align}
g_A&=\bra{A_1}H_0\ket{t_Q}=\ii t_2(1-e^{-\ii k})u_A,\nonumber\\
g_B&=\bra{B_1}H_0\ket{t_Q}=t_1u_A-\ii t_2(1-e^{-\ii k})u_B,
\label{eq:haldaneg}
\end{align}
with $G^2=|g_A|^2+|g_B|^2$. Normalization is over the entire inner-edge wave function, so $|u_A|^2+|u_B|^2$ is its weight on the first inner cell. The boundary basis is $B=(g_A,g_B)^T/G$ and $D=(-g_B^*,g_A^*)^T/G$. Thus, one retained edge state selects one directly coupled bright state even for rank-two intercell hopping.

In the $(B,D,t_Q)$ basis, the effective Hamiltonian is
\begingroup
\setlength{\arraycolsep}{4pt}
\begin{equation}
H_{\rm bd}^{\rm H}=
\begin{pmatrix}
h_{BB}-\ii\gamma+\Sigma^{\rm reg}_{BB}&h_{BD}+\Sigma^{\rm reg}_{BD}&G\\
h_{DB}+\Sigma^{\rm reg}_{DB}&h_{DD}-\ii\gamma+\Sigma^{\rm reg}_{DD}&0\\
G&0&\epsilon_Q
\end{pmatrix}.
\label{eq:haldanebd}
\end{equation}
\endgroup
Here $h_{mn}$ are the projected outer-cell matrix elements defined in Eq.~\eqref{eq:hdefinitions}. Appendix~\ref{app:haldane} gives the hopping blocks and the pole-subtracted surface self-energy. Both $|B\rangle$ and $|D\rangle$ carry the boundary loss $-\ii\gamma$. The dark state has no direct coupling to $|t_Q\rangle$ but can couple to $|B\rangle$ through the outer Hamiltonian and the off-diagonal self-energy. At sufficiently large $\gamma$, the reconstructed mode has bright amplitude of order $G/\gamma$ and therefore $\Gamma_{\rm in}\simeq G^2/\gamma$. The matrix-valued self-energy determines corrections to the leading rate and the poles at finite $\gamma$. Figure~\ref{fig:haldane}(b) shows the eigenvalues $\lambda_\pm$ of $T^\dagger T$. Unlike $\lambda_\pm$, $G^2$ also depends on the normalized inner-edge wave function. In the QWZ model, rank-one hopping confines the residual self-energy to $BB$. In the Haldane model, the nonzero $DD$ and $BD$ entries in Fig.~\ref{fig:haldane}(c) reveal additional return paths through the inner system. A single isolated bright--inner dimer therefore omits physical processes at generic momentum.

At $k=\pi$, symmetry gives $h_{\rm out}=0$, $\epsilon_Q=0$, and $\Sigma^{\rm reg}_{BD}=\Sigma^{\rm reg}_{DB}=0$. The bright--inner pair then obeys
\begin{equation}
\mathcal F(E,\gamma)=E[E+\ii\gamma-\Sigma^{\rm reg}_{BB}(E,\pi)]-G^2=0.
\label{eq:haldaneP}
\end{equation}
The lossy dark channel obeys the separate equation $E+\ii\gamma-\Sigma^{\rm reg}_{DD}(E,\pi)=0$. Unlike the resonant QWZ dimer, the Haldane pair retains a nonzero residual self-energy at this symmetry point. Evaluating the self-energy at a physical pole energy, we obtain the exact equation derived in Appendix~\ref{app:haldane}
\begin{equation}
E^2+\ii\left(\gamma-\frac{\lambda_-}{\gamma}\right)E-G^2=0,
\qquad G^2=\lambda_+-\lambda_- ,
\label{eq:haldanequadratic}
\end{equation}
where $\lambda_\pm$ are evaluated at $k=\pi$. The two decaying localized roots exist for $\gamma>\sqrt{\lambda_-}$ and coalesce at $\gamma_{\rm EP}=G+\sqrt{\lambda_+}=2.27\,t_1$ and $E_{\rm EP}=-\ii G=-1.132\,\ii t_1$. The pole solutions reproduce the pair in Fig.~\ref{fig:haldane}(d). Omitting the residual self-energy would instead give $\gamma_{\rm EP}=2G=2.263\,t_1$. The remaining inner states shift the EP through the contribution $\lambda_-/\gamma$. 

\section{Discussion and conclusions}
\label{sec:conclusion}
Boundary loss changes both where a chiral edge mode resides and how it decays. Weak loss acts on the original edge through its boundary weight. Strong loss, however, suppresses transfer into the outer layer, leaving an inner-edge mode. Its residual decay arises from coupling with a lossy bright state, while its underlying Hermitian branch is topologically protected. This coupling to the dissipative edge state ultimately determines its decay.

The unitary transformation $U(k)$ distinguishes these mechanisms even when their limiting rates share the same momentum-dependent edge weight; the matrix elements in Figs.~\ref{fig:qwzresults}(e),(f) reveal information beyond decay rate. Haldane shares the leading strong-loss mechanism, but feedback from the remaining inner states enters through a matrix self-energy and shifts the EP away from the isolated-dimer prediction. The inverse-loss asymptote and EP coalescence thus characterize distinct aspects of the effective model.

The asymptotic laws above apply to uniform pure loss with an isolated edge branch and should not be generalized to arbitrary boundary non-Hermiticity. Taking the QWZ model as an example, outer-row nonreciprocal hopping neither induces boundary reconstruction nor reproduces the same inverse-loss law. Balanced gain and loss on the two outer orbitals, by contrast, can support a reconstructed real-energy branch. These operator-specific examples demonstrate that the projection framework is reusable, while the decay and reconstruction behaviors depend on the specific boundary operator.

Following the frequency-resolved field mapping of
~\cite{PhysRevLett.132.113802}, joint
measurements of transverse mode profiles, resonance
frequencies, and linewidths could test our effective
model. The inward shift together with the change
from $\Gamma\simeq\gamma Z_k$ to
$\Gamma\simeq G_k^2/\gamma$ would test the transition
from direct edge damping to bright-state-mediated
decay. This protocol is applicable to topolectrical
circuits, transmission-line networks, and phononic
crystals supporting chiral edge modes with tunable
boundary loss. Moreover, another natural extension is spatially nonuniform boundary
loss, such as $\gamma(x)=\gamma_0+\gamma_1\cos(Qx+\phi)$
with $\gamma_0\geq|\gamma_1|$. Such modulation couples
different edge momenta, raising the question of how
chiral propagation and spatially varying boundary
reconstruction jointly determine the profiles and
lifetimes of long-lived modes.

\begin{acknowledgments}
We are grateful to Xue-Ping Ren for helpful discussions. This work was supported by the National Key R\&D Program of China (Grant No.~2023YFA1406704). Q. Du acknowledges support from the Natural Science Foundation of Shandong Province of China (Grant No.~ZR2024QA141) and the Research Start-up Foundation for High-level Talents of Linyi University (Grant No.~Z6125007).
\end{acknowledgments}

\appendix
\section{Boundary projection and induced decay}
\label{app:general}
Firstly, we separate the lossy layer from the Hermitian inner system. At fixed momentum, let $x$ and $\psi$ denote outer- and inner-layer amplitudes. The hopping $\mathcal T$ maps inner to outer layers, with $T_{\rm out,in}=\langle1|H_0|2\rangle$ for nearest-layer hopping:

\begin{equation}
\begin{pmatrix}
h_{\rm out}-\ii\gamma I&\mathcal T\\
\mathcal T^\dagger&H_{\rm in}
\end{pmatrix}
\begin{pmatrix}x\\\psi\end{pmatrix}
=E\begin{pmatrix}x\\\psi\end{pmatrix}.
\label{eq:Ablock}
\end{equation}
For a normalized, isolated state $H_{\rm in}\ket{t_Q}=\epsilon_Q\ket{t_Q}$ of the Hermitian inner system, write
\begin{equation}
\begin{gathered}
\psi=c_Q\ket{t_Q}+\chi,\qquad
Q_{\rm in}=I_{\rm in}-\ket{t_Q}\bra{t_Q},\\
Q_{\rm in}\chi=\chi,\qquad
\braket{t_Q|\chi}=0.
\end{gathered}
\label{eq:Asplit}
\end{equation}
With $g=\mathcal T\ket{t_Q}$ and $H_r=Q_{\rm in}H_{\rm in}Q_{\rm in}$ on $\operatorname{ran}Q_{\rm in}$, projection gives
\begin{align}
(E-h_{\rm out}+\ii\gamma I)x&=g c_Q+\mathcal T\chi,\nonumber\\
(E-\epsilon_Q)c_Q&=g^\dagger x,\nonumber\\
(E-H_r)\chi&=Q_{\rm in}\mathcal T^\dagger x.
\label{eq:Aamplitudes}
\end{align}
The selected edge state communicates with the remaining inner states only through the outer layer.

Secondly, we eliminate the other inner states. Eliminating $\chi$ gives the residual self-energy (where the inverse exists):
\begin{align}
\chi&=(E-H_r)^{-1}Q_{\rm in}\mathcal T^\dagger x,\nonumber\\
\Sigma^{\rm reg}(E)&=\mathcal TQ_{\rm in}(E-H_r)^{-1}
Q_{\rm in}\mathcal T^\dagger.
\label{eq:Aeliminate}
\end{align}
For inner eigenstates $\ket\nu$ with energies $\epsilon_\nu$, where $r$ excludes $t_Q$, equivalently
\begin{equation}
\Sigma^{\rm reg}(E)=\sum_{\nu\in r}
\frac{v_\nu v_\nu^\dagger}{E-\epsilon_\nu},
\qquad v_\nu=\mathcal T\ket\nu.
\label{eq:Aspectral}
\end{equation}
Subtracting the retained edge pole from the full surface feedback gives
\begin{equation}
\mathcal T(E-H_{\rm in})^{-1}\mathcal T^\dagger
=\frac{gg^\dagger}{E-\epsilon_Q}+\Sigma^{\rm reg}(E),
\label{eq:Apolesubtract}
\end{equation}
and the effective equation
\begin{equation}
\begin{pmatrix}
h_{\rm out}-\ii\gamma I+\Sigma^{\rm reg}(E)&g\\
g^\dagger&\epsilon_Q
\end{pmatrix}
\begin{pmatrix}x\\c_Q\end{pmatrix}
=E\begin{pmatrix}x\\c_Q\end{pmatrix}.
\label{eq:Aretained}
\end{equation}
The pole subtraction avoids double counting. The remaining self-energy has no explicit $\gamma$ dependence and acts only on the outer block.

Thirdly, we construct the bright and dark directions. The coupling vector $g$ selects a single bright outer channel. For $G=\sqrt{g^\dagger g}>0$, let $B=g/G$ and $R=(B,D_1,\ldots,D_{N_s-1})$:

\begin{equation}
R^\dagger g=(G,0,\ldots,0)^T,\qquad
x=R(x_B,x_D)^T.
\label{eq:Arotation}
\end{equation}
With $h=R^\dagger h_{\rm out}R$ and $\widetilde h=h+R^\dagger\Sigma^{\rm reg}R$, the dark amplitude and inner-edge pole satisfy
\begin{equation}
x_D=(E-\widetilde h_{DD}+\ii\gamma I_D)^{-1}
\widetilde h_{DB}x_B.
\label{eq:Adarkamplitude}
\end{equation}
\begin{equation}
\begin{gathered}
E-\epsilon_Q=\frac{G^2}{E-\widetilde h_{BB}+\ii\gamma-\Lambda_D(E)},\\
\Lambda_D(E)=\widetilde h_{BD}
(E-\widetilde h_{DD}+\ii\gamma I_D)^{-1}\widetilde h_{DB}.
\end{gathered}
\label{eq:Adarkfeedback}
\end{equation}
Only $B$ couples directly to $t_Q$; dark channels can still shift its pole through feedback. If $G=0$, the inner state is decoupled from the outer layer.

Fourthly, we reconstruct the physical state and its decay rate. The eliminated inner amplitudes contribute to the full norm. The physical eigenstate and decay rate are
\begin{equation}
\ket\Phi=\begin{pmatrix}x\\c_Q\ket{t_Q}+\chi\end{pmatrix},
\qquad
\|\Phi\|^2=\|x\|^2+|c_Q|^2+\|\chi\|^2,
\label{eq:Afullstate}
\end{equation}
\begin{equation}
\Gamma\equiv-\operatorname{Im}E
=\frac{\gamma\|x\|^2}{\|x\|^2+|c_Q|^2+\|\chi\|^2}.
\label{eq:Anormrate}
\end{equation}
Thus decay measures the probability on the lossy layer.

At large $\gamma$, eliminating the outer layer gives
\begin{align}
H_{\rm in}^{\rm eff}(E)
&=H_{\rm in}+\mathcal T^\dagger
(E-h_{\rm out}+\ii\gamma I)^{-1}\mathcal T\nonumber\\
&=H_{\rm in}-\frac{\ii}{\gamma}K
+\frac{1}{\gamma^2}L(E)+O(\gamma^{-3}),\nonumber\\
K&=\mathcal T^\dagger\mathcal T,\qquad
L(E)=\mathcal T^\dagger(E-h_{\rm out})\mathcal T,\nonumber\\
\Gamma_{\rm in}&=\frac{\bra{t_Q}K\ket{t_Q}}{\gamma}
+O(\gamma^{-3})
=\frac{G^2}{\gamma}+O(\gamma^{-3}).
\label{eq:Alargeloss}
\end{align}
For an isolated, nondegenerate inner state, the next energy correction is
\begin{align}
E={}&\epsilon_Q-\frac{\ii G^2}{\gamma}
+\frac{1}{\gamma^2}\left[
\bra{t_Q}L(\epsilon_Q)\ket{t_Q}\right.\nonumber\\
&\left.\hspace{12mm}-\sum_{\nu\in r}
\frac{|\bra\nu K\ket{t_Q}|^2}{\epsilon_Q-\epsilon_\nu}
\right]+O(\gamma^{-3}).
\label{eq:Aenergyexpansion}
\end{align}
Strong loss suppresses the outer weight to order $G^2/\gamma^2$, yielding $\Gamma_{\rm in}\simeq G^2/\gamma$. The order-$\gamma^{-2}$ energy shift is real. These expansions require sufficiently large $\gamma$ and a spectrally isolated inner state.

To locate an EP, retain the energy dependence of the self-energy and define
\begin{equation}
F(E,\gamma)=\det[E I-H_{\rm bd}(E,\gamma)].
\label{eq:Asecular}
\end{equation}
A second-order EP requires $F=\partial_EF=0$, $\partial_E^2F\neq0$, and a one-dimensional eigenspace; $\partial_E$ acts on $\Sigma^{\rm reg}$ as well. Only an isolated resonant bright--inner dimer reduces to $\gamma_{\rm EP}=2G$.

\section{QWZ effective Hamiltonian}
\label{app:qwz}

For $k\equiv k_x$, set $a_k=t\sin k$ and $b_k=M+t\cos k$. With $T=\langle y|H_0|y+1\rangle$, the strip Hamiltonian reads Eq.~\eqref{eq:qwzstrip}. Here $\sigma_i$ act on the two orbitals of each row. In the $\sigma_x$ basis,
\begin{equation}
\ket{+}_x=\frac{1}{\sqrt{2}}
\begin{pmatrix}1\\1\end{pmatrix},
\qquad
\ket{-}_x=\frac{1}{\sqrt{2}}
\begin{pmatrix}1\\-1\end{pmatrix}.
\label{eq:Bsigmaxbasis}
\end{equation}
\begin{equation}
h_x(k)=
\begin{pmatrix}
a_k&b_k\\
b_k&-a_k
\end{pmatrix},
\qquad
T_x=
\begin{pmatrix}
0&t\\
0&0
\end{pmatrix}
=t\ket{+}_x{}_x\!\bra{-}.
\label{eq:Bblocks}
\end{equation}
The transverse chain is $\ket{1,-}_x
\overset{b_k}{\longleftrightarrow}
\ket{1,+}_x
\overset{t}{\longleftrightarrow}
\ket{2,-}_x
\overset{b_k}{\longleftrightarrow}
\ket{2,+}_x
\overset{t}{\longleftrightarrow}
\ket{3,-}_x
\cdots .$
For $N_y\to\infty$, let $r_k=-b_k/t$ and $Z_k=1-r_k^2$. The normalized inner-edge state ($|r_k|<1$) is
\begin{equation}
\ket{t_Q}=\sqrt{Z_k}\sum_{n=0}^{\infty}r_k^n\ket{2+n,-}_x,
\qquad \epsilon_Q=-a_k.
\label{eq:Bwaves}
\end{equation}
Its localization length obeys $\xi_y^{-1}=-\ln|r_k|$. Rank-one hopping selects $B=\ket{1,+}_x$ with $G_k=t\sqrt{Z_k}$; the edge merges with the bulk at $|r_k|=1$.
Let $G_s=[EI-h_x-T_xG_sT_x^\dagger]^{-1}$ be the Hermitian inner surface resolvent. And $\eta$ denotes the scalar surface self-energy obtained by projecting $T_x G_s T_x^\dagger$ onto the bright-state direction.
\begin{equation}
\eta=\frac{t^2(E-a_k-\eta)}{(E+a_k)(E-a_k-\eta)-b_k^2},
\label{eq:Bsurfaceratio}
\end{equation}
\begin{equation}
(E+a_k)\eta^2-(E^2-a_k^2-b_k^2+t^2)\eta+t^2(E-a_k)=0.
\label{eq:Bsurfacequadratic}
\end{equation}
On the physical branch ($\eta\sim t^2/E$ at large $|E|$), subtract the inner-edge pole:

\begin{equation}
\eta(E,k)=\frac{G_k^2}{E+a_k}+\Sigma^{\mathrm{reg}}_{BB}(E,k).
\label{eq:Bpole}
\end{equation}
The boundary spectrum follows from
\begin{equation}
(E+a_k+\ii\gamma)(E-a_k+\ii\gamma-\eta)-b_k^2=0,
\label{eq:Bsecular}
\end{equation}
using the physical solution of Eq.~\eqref{eq:Bsurfacequadratic}.
The original Hermitian edge $t_0$ and its orthogonal companion $n_-$ obey
\begin{equation}
\ket{t_0}=\sqrt{Z_k}\ket D+r_k\ket{t_Q},\qquad
\ket{n_-}=r_k\ket D-\sqrt{Z_k}\ket{t_Q}.
\label{eq:Brotation}
\end{equation}
\begin{gather}
\bra{n_-}H_0\ket B=r_kb_k-\sqrt{Z_k}G_k=-t,\nonumber\\
\bra{t_0}H_0\ket B=\sqrt{Z_k}b_k+r_kG_k=0,\nonumber\\
U^\dagger P_1U=\begin{pmatrix}
r_k^2&0&r_k\sqrt{Z_k}\\0&1&0\\r_k\sqrt{Z_k}&0&Z_k
\end{pmatrix}.
\label{eq:Bweightmatrix}
\end{gather}
The rotation separates direct boundary loss from dissipative mixing. Eliminating $(n_-,B)$ in the weak-loss basis gives
\begin{align}
E+a_k+\ii\gamma Z_k&=\Sigma_t^{\rm mix}(E,\gamma),\nonumber\\
\Sigma_t^{\rm mix}(E,\gamma)
&=-\gamma^2r_k^2Z_k\,
\frac{C(E,\gamma)}
{(E+a_k+\ii\gamma r_k^2)C(E,\gamma)-t^2},\nonumber\\
C(E,\gamma)&=E-a_k+\ii\gamma-\Sigma^{\mathrm{Reg}}(E,k).
\label{eq:Bweakscalar}
\end{align}
For an isolated Hermitian edge at $E_0=-a_k$, the regular self-energy coefficients are
\begin{align}
\Sigma^{\mathrm{Reg}}(-a_k,k)&=\frac{2a_kr_k^2}{Z_k},\nonumber\\
\left.\partial_E\Sigma^{\mathrm{Reg}}(E,k)\right|_{E=-a_k}
&=-\frac{r_k^2}{Z_k}
-\frac{4a_k^2r_k^2}{t^2Z_k^3},
\label{eq:Bregderivatives}
\end{align}
and expansion at small $\gamma$ yields
\begin{equation}
E_{\rm weak}=-a_k-\ii\gamma Z_k
-\frac{2a_kr_k^2}{t^2}\gamma^2
-\ii\frac{4a_k^2r_k^2}{t^4}\gamma^3+O(\gamma^4).
\label{eq:Bweakexpansion}
\end{equation}
The $\gamma^2$ term shifts only the real energy; mixing first changes the decay at order $\gamma^3$. This expansion is not uniform near $|r_k|=1$.

At $M=-t$ and $k=0$, the bright--inner dimer is exact. For its normalized right eigenvectors $\widehat v_\pm$, let $V=(\widehat v_+,\widehat v_-)$. Then
\begin{equation}
\sigma=s_{\min}(V)=
\begin{cases}
\sqrt{1-\gamma/(2t)},&0\leq\gamma\leq2t,\\
\sqrt{1-2t/\gamma},&\gamma\geq2t.
\end{cases}
\label{eq:Bsigma}
\end{equation}
The vanishing singular value at $\gamma=2t$ signals eigenvector coalescence; the original dark edge state remains separate.

\section{Haldane effective Hamiltonian}
\label{app:haldane}

For zigzag edges with $M=0$ and $\phi=\pi/2$, the $(A,B)$-sublattice blocks ($T=\langle y|H_0|y+1\rangle$) are
\begin{equation}
\begin{gathered}
h(k)=\begin{pmatrix}
2t_2\sin k&t_1(1+e^{-\ii k})\\
t_1(1+e^{\ii k})&-2t_2\sin k
\end{pmatrix},\\[2pt]
T(k)=\begin{pmatrix}
\ii t_2(1-e^{-\ii k})&0\\
t_1&-\ii t_2(1-e^{-\ii k})
\end{pmatrix}.
\end{gathered}
\label{eq:Cblocks}
\end{equation}
For the first inner-cell spinor $u_Q=(u_A,u_B)^T$ of the globally normalized edge state, let $v=Tu_Q=(g_A,g_B)^T$ and $G=\sqrt{v^\dagger v}$, $B=v/G$, $D=(-B_2^*,B_1^*)^T$, $R=(B,D)$. The pole-subtracted Hermitian surface feedback is
\begin{align}
G_s&=[EI_2-h-TG_sT^\dagger]^{-1},\nonumber\\
\Sigma_{\rm bd}^{\rm reg}(E,k)
&=R^\dagger\left[TG_sT^\dagger-\frac{vv^\dagger}{E-\epsilon_Q}\right]R,
\label{eq:Csubtract}
\end{align}
where the physical branch has $G_s\sim I_2/E$ for large $|E|$. The subtracted pole is the retained inner-edge state. The two eigenvalues of $T^\dagger T$ are
\begin{align}
\lambda_\pm(k)={}&4t_2^2\sin^2(k/2)+\frac{t_1^2}{2}\nonumber\\
&\pm\frac12\sqrt{t_1^4+16t_1^2t_2^2\sin^2(k/2)}.
\label{eq:Clambdas}
\end{align}

At $k=\pi$, set $p=\sqrt{\lambda_+}$ and $q=\sqrt{\lambda_-}$ ($p>q$). Here $h=0$ and
\begin{equation}
R^\dagger TR=\begin{pmatrix}0&-p\\q&0\end{pmatrix}.
\label{eq:CrotatedT}
\end{equation}
The zero-energy inner edge occupies the $D$ direction on $y=2,4,\ldots$. Normalization yields
\begin{equation}
\begin{aligned}
f_{2m+2}&=\frac{q}{p}f_{2m},\qquad
|f_2|^2=1-\frac{q^2}{p^2},\\
G^2&=p^2|f_2|^2=p^2-q^2.
\end{aligned}
\label{eq:Cweight}
\end{equation}
At this momentum, surface feedback reduces to
\begin{equation}
\Sigma_B^{\rm full}=\frac{p^2}{E-\Sigma_D^{\rm full}},\qquad
\Sigma_D^{\rm full}=\frac{q^2}{E-\Sigma_B^{\rm full}}.
\label{eq:Cscalarpair}
\end{equation}
\begin{equation}
\Sigma_B^{\rm full}-\Sigma_D^{\rm full}=\frac{G^2}{E},
\qquad \Sigma_{BB}^{\rm reg}=\Sigma_D^{\rm full}.
\label{eq:Cpole_subtraction}
\end{equation}
With the bright--inner pole condition $E+\ii\gamma=\Sigma_B^{\rm full}$, its on-shell feedback and EP satisfy
\begin{equation}
\Sigma_{BB}^{\rm reg}=\frac{q^2}{-\ii\gamma}
=\frac{\ii q^2}{\gamma}.
\label{eq:Con_shell_feedback}
\end{equation}
Substituting this into the bright–inner pole equation yields
\begin{equation}
\begin{aligned}
E^2+\ii\left(\gamma-\frac{q^2}{\gamma}\right)E-G^2&=0,\\
\gamma_{\rm EP}=G+p,\qquad E_{\rm EP}&=-\ii G.
\end{aligned}
\label{eq:Croots}
\end{equation}
The on-shell relation does not make the self-energy energy independent away from the pole. The two-cell localization ratio is
\begin{equation}
z=\frac{q}{p}\left(1-\frac{\ii E}{\gamma}\right).
\label{eq:Clocalization}
\end{equation}
The localized pair lies on the physical half-space branch for $\gamma>q$. At $t_2/t_1=0.2$, $\gamma_{\rm EP}/t_1=2.272$; residual feedback shifts the EP, while the strong-loss rate still scales as $G^2/\gamma$.

\makeatletter
\immediate\write\@auxout{\string\citation{PRBTitleControl}}
\makeatother
\bibliographystyle{apsrev4-2}
\bibliography{refs}

@book{Breuer_oqs,
    author = {Breuer, Heinz-Peter and Petruccione, Francesco},
    title = {The Theory of Open Quantum Systems},
    publisher = {Oxford University Press},
    year = {2007},
    month = {01},
    isbn = {9780199213900},
    doi = {10.1093/acprof:oso/9780199213900.001.0001},
    url = {https://doi.org/10.1093/acprof:oso/9780199213900.001.0001},
    address={New York}
}

@article{PhysRevB.102.115109,
  title = {Universal Lindblad equation for open quantum systems},
  author = {Nathan, Frederik and Rudner, Mark S.},
  journal = {Phys. Rev. B},
  volume = {102},
  issue = {11},
  pages = {115109},
  numpages = {24},
  year = {2020},
  month = {Sep},
  publisher = {American Physical Society},
  doi = {10.1103/PhysRevB.102.115109},
  url = {https://link.aps.org/doi/10.1103/PhysRevB.102.115109}
}

@article{RevModPhys.94.045006,
  title = {Nonequilibrium boundary-driven quantum systems: Models, methods, and properties},
  author = {Landi, Gabriel T. and Poletti, Dario and Schaller, Gernot},
  journal = {Rev. Mod. Phys.},
  volume = {94},
  issue = {4},
  pages = {045006},
  numpages = {58},
  year = {2022},
  month = {Dec},
  publisher = {American Physical Society},
  doi = {10.1103/RevModPhys.94.045006},
  url = {https://link.aps.org/doi/10.1103/RevModPhys.94.045006}
}

@article{RevModPhys.93.015005,
  title = {Exceptional topology of non-Hermitian systems},
  author = {Bergholtz, Emil J. and Budich, Jan Carl and Kunst, Flore K.},
  journal = {Rev. Mod. Phys.},
  volume = {93},
  issue = {1},
  pages = {015005},
  numpages = {31},
  year = {2021},
  month = {Feb},
  publisher = {American Physical Society},
  doi = {10.1103/RevModPhys.93.015005},
  url = {https://link.aps.org/doi/10.1103/RevModPhys.93.015005}
}

@article{Heiss_2012,
doi = {10.1088/1751-8113/45/44/444016},
url = {https://dx.doi.org/10.1088/1751-8113/45/44/444016},
year = {2012},
month = {oct},
publisher = {IOP Publishing},
volume = {45},
number = {44},
pages = {444016},
author = {Heiss, W D},
title = {The physics of exceptional points},
journal = {J. Phys. A: Math. Theor.},
}

@article{RevModPhys.82.3045,
  title = {Colloquium: Topological insulators},
  author = {Hasan, M. Z. and Kane, C. L.},
  journal = {Rev. Mod. Phys.},
  volume = {82},
  issue = {4},
  pages = {3045--3067},
  numpages = {0},
  year = {2010},
  month = {Nov},
  publisher = {American Physical Society},
  doi = {10.1103/RevModPhys.82.3045},
  url = {https://link.aps.org/doi/10.1103/RevModPhys.82.3045}
}

@article{Zou2021,
  author    = {Deyuan Zou and Tian Chen and Wenjing He and Jiacheng Bao and Ching Hua Lee and Houjun Sun and Xiangdong Zhang},
  title     = {Observation of hybrid higher-order skin-topological effect in non-Hermitian topolectrical circuits},
  journal   = {Nat. Commun.},
  year      = {2021},
  volume    = {12},
  number    = {1},
  pages     = {7201},
  month     = {dec},
  doi       = {10.1038/s41467-021-26414-5},
  issn      = {2041-1723},
  url       = {https://doi.org/10.1038/s41467-021-26414-5},
}

@article{PhysRevLett.123.073601,
  title = {Higher-Order Topological Corner States Induced by Gain and Loss},
  author = {Luo, Xi-Wang and Zhang, Chuanwei},
  journal = {Phys. Rev. Lett.},
  volume = {123},
  issue = {7},
  pages = {073601},
  numpages = {8},
  year = {2019},
  month = {Aug},
  publisher = {American Physical Society},
  doi = {10.1103/PhysRevLett.123.073601},
  url = {https://link.aps.org/doi/10.1103/PhysRevLett.123.073601}
}

@article{PhysRevLett.128.223903,
  title = {Gain-Loss-Induced Hybrid Skin-Topological Effect},
  author = {Li, Yaohua and Liang, Chao and Wang, Chenyang and Lu, Cuicui and Liu, Yong-Chun},
  journal = {Phys. Rev. Lett.},
  volume = {128},
  issue = {22},
  pages = {223903},
  numpages = {7},
  year = {2022},
  month = {Jun},
  publisher = {American Physical Society},
  doi = {10.1103/PhysRevLett.128.223903},
  url = {https://link.aps.org/doi/10.1103/PhysRevLett.128.223903}
}

@article{PhysRevB.106.035425,
  title = {Hybrid skin-topological modes without asymmetric couplings},
  author = {Zhu, Weiwei and Gong, Jiangbin},
  journal = {Phys. Rev. B},
  volume = {106},
  issue = {3},
  pages = {035425},
  numpages = {13},
  year = {2022},
  month = {Jul},
  publisher = {American Physical Society},
  doi = {10.1103/PhysRevB.106.035425},
  url = {https://link.aps.org/doi/10.1103/PhysRevB.106.035425}
}

@article{PhysRevResearch.6.013213,
  title = {Non-Hermitian chiral skin effect},
  author = {Ma, Xin-Ran and Cao, Kui and Wang, Xiao-Ran and Wei, Zheng and Du, Qian and Kou, Su-Peng},
  journal = {Phys. Rev. Res.},
  volume = {6},
  issue = {1},
  pages = {013213},
  numpages = {17},
  year = {2024},
  month = {Feb},
  publisher = {American Physical Society},
  doi = {10.1103/PhysRevResearch.6.013213},
  url = {https://link.aps.org/doi/10.1103/PhysRevResearch.6.013213}
}

@article{PRXQuantum.4.030315,
  title = {Hermitian Bulk  --  Non-Hermitian Boundary Correspondence},
  author = {Schindler, Frank and Gu, Kaiyuan and Lian, Biao and Kawabata, Kohei},
  journal = {PRX Quantum},
  volume = {4},
  issue = {3},
  pages = {030315},
  numpages = {18},
  year = {2023},
  month = {Aug},
  publisher = {American Physical Society},
  doi = {10.1103/PRXQuantum.4.030315},
  url = {https://link.aps.org/doi/10.1103/PRXQuantum.4.030315}
}

@article{PhysRevLett.131.256602,
  title = {Universal Platform of Point-Gap Topological Phases from Topological Materials},
  author = {Nakamura, D. and Inaka, Kazuya and Okuma, Nobuyuki and Sato, Masatoshi},
  journal = {Phys. Rev. Lett.},
  volume = {131},
  issue = {25},
  pages = {256602},
  numpages = {6},
  year = {2023},
  month = {Dec},
  publisher = {American Physical Society},
  doi = {10.1103/PhysRevLett.131.256602},
  url = {https://link.aps.org/doi/10.1103/PhysRevLett.131.256602}
}

@article{PhysRevLett.132.113802,
  title = {Localization of Chiral Edge States by the Non-Hermitian Skin Effect},
  author = {Liu, Gui-Geng and Mandal, Subhaskar and Zhou, Peiheng and Xi, Xiang and Banerjee, Rimi and Hu, Yuan-Hang and Wei, Minggui and Wang, Maoren and Wang, Qiang and Gao, Zhen and Chen, Hongsheng and Yang, Yihao and Chong, Yidong and Zhang, Baile},
  journal = {Phys. Rev. Lett.},
  volume = {132},
  issue = {11},
  pages = {113802},
  numpages = {6},
  year = {2024},
  month = {Mar},
  publisher = {American Physical Society},
  doi = {10.1103/PhysRevLett.132.113802},
  url = {https://link.aps.org/doi/10.1103/PhysRevLett.132.113802}
}

@article{Zhao_2019,
author = {Han Zhao  and Xingdu Qiao  and Tianwei Wu  and Bikashkali Midya  and Stefano Longhi  and Liang Feng },
title = {Non-Hermitian topological light steering},
journal = {Science},
volume = {365},
number = {6458},
pages = {1163-1166},
year = {2019},
doi = {10.1126/science.aay1064},
url = {https://www.science.org/doi/abs/10.1126/science.aay1064}}

@article{PhysRevLett.129.053903,
  title = {Effective Hamiltonian for Photonic Topological Insulator with Non-Hermitian Domain Walls},
  author = {Li, Yandong and Fan, Chongxiao and Hu, Xiaoyong and Ao, Yutian and Lu, Cuicui and Chan, C. T. and Kennes, Dante M. and Gong, Qihuang},
  journal = {Phys. Rev. Lett.},
  volume = {129},
  issue = {5},
  pages = {053903},
  numpages = {6},
  year = {2022},
  month = {Jul},
  publisher = {American Physical Society},
  doi = {10.1103/PhysRevLett.129.053903},
  url = {https://link.aps.org/doi/10.1103/PhysRevLett.129.053903}
}

@article{Wu2023,
  author    = {Deguang Wu and Jiasong Chen and Wei Su and Rui Wang and Baigeng Wang and D. Y. Xing},
  title     = {Effective impurity behavior emergent from non-Hermitian proximity effect},
  journal   = {Commun. Phys.},
  year      = {2023},
  volume    = {6},
  number    = {1},
  pages     = {160},
  month     = {jul},
  doi       = {10.1038/s42005-023-01282-1},
  issn      = {2399-3650},
  url       = {https://doi.org/10.1038/s42005-023-01282-1},
}

@article{Du2024,
  author    = {Qian Du and Xin-Ran Ma and Su-Peng Kou},
  title     = {Non-Hermitian tearing by dissipation},
  journal   = {Eur. Phys. J. B},
  year      = {2024},
  volume    = {97},
  number    = {6},
  pages     = {85},
  month     = {jun},
  doi       = {10.1140/epjb/s10051-024-00714-3},
  issn      = {1434-6036},
  url       = {https://doi.org/10.1140/epjb/s10051-024-00714-3},
}

@article{QWZ2006,
  author = {Qi, Xiao-Liang and Wu, Yong-Shi and Zhang, Shou-Cheng},
  title = {Topological quantization of the spin {Hall} effect in two-dimensional paramagnetic semiconductors},
  journal = {Phys. Rev. B},
  volume = {74},
  pages = {085308},
  year = {2006},
  doi = {10.1103/PhysRevB.74.085308}
}

@article{Haldane1988,
  author = {Haldane, F. D. M.},
  title = {Model for a Quantum {Hall} Effect without {Landau} Levels: Condensed-Matter Realization of the {Parity Anomaly}},
  journal = {Phys. Rev. Lett.},
  volume = {61},
  pages = {2015--2018},
  year = {1988},
  doi = {10.1103/PhysRevLett.61.2015}
}

@article{Feshbach1958,
  author = {Feshbach, Herman},
  title = {Unified theory of nuclear reactions},
  journal = {Ann. Phys.},
  volume = {5},
  pages = {357--390},
  year = {1958},
  doi = {10.1016/0003-4916(58)90007-1}
}

@article{Lu2024,
  author = {Lu, Ming and Liu, Xue-Zhu and Li, Hailong and Zhang, Zhi-Qiang and Liu, Jie and Xie, X. C.},
  title = {Dissipation-enhanced unidirectional transport in topological systems},
  journal = {Phys. Rev. B},
  volume = {109},
  pages = {165402},
  year = {2024},
  doi = {10.1103/PhysRevB.109.165402}
}

@CONTROL{PRBTitleControl, title="0"}

@article{PhysRevLett.131.083801,
  author = {Wetter, Helene and Fleischhauer, Michael and Linden, Stefan and Schmitt, Julian},
  title = {Observation of a Topological Edge State Stabilized by Dissipation},
  journal = {Phys. Rev. Lett.},
  volume = {131},
  pages = {083801},
  year = {2023},
  doi = {10.1103/PhysRevLett.131.083801}
}

@article{PhysRevResearch.5.043229,
  author = {Yang, Fan and Molignini, Paolo and Bergholtz, Emil J.},
  title = {Dissipative boundary state preparation},
  journal = {Phys. Rev. Research},
  volume = {5},
  pages = {043229},
  year = {2023},
  doi = {10.1103/PhysRevResearch.5.043229}
}

@article{PhysRevLett.131.256601,
  author = {Hegde, Suraj S. and Ehmcke, Toni and Meng, Tobias},
  title = {Edge-Selective Extremal Damping from Topological Heritage of Dissipative {Chern} Insulators},
  journal = {Phys. Rev. Lett.},
  volume = {131},
  pages = {256601},
  year = {2023},
  doi = {10.1103/PhysRevLett.131.256601}
}

@article{PhysRevLett.135.266602,
  author = {Zhou, Yan-Chen and Lai, Hua-Shan and Sun, Ze-Qun and Sun, Xiao-Chen and Xie, Jian-Lan and He, Cheng and Chen, Yan-Feng},
  title = {Loss-Induced Bulk-Boundary Detachment in a Photonic {Chern} Insulator},
  journal = {Phys. Rev. Lett.},
  volume = {135},
  pages = {266602},
  year = {2025},
  doi = {10.1103/6lyv-yrq2}
}

@article{R-TKNN,
  author = {Thouless, D. J. and Kohmoto, M. and Nightingale, M. P. and den Nijs, M.},
  title = {Quantized {Hall} Conductance in a Two-Dimensional Periodic Potential},
  journal = {Phys. Rev. Lett.},
  volume = {49},
  pages = {405--408},
  year = {1982},
  doi = {10.1103/PhysRevLett.49.405},
  url = {https://journals.aps.org/prl/abstract/10.1103/PhysRevLett.49.405}
}

@article{Verstraete2009,
  author = {Verstraete, Frank and Wolf, Michael M. and Cirac, J. Ignacio},
  title = {Quantum computation and quantum-state engineering driven by dissipation},
  journal = {Nat. Phys.},
  volume = {5},
  pages = {633--636},
  year = {2009},
  doi = {10.1038/nphys1342},
  url = {https://www.nature.com/articles/nphys1342}
}

@article{Diehl2011,
  author = {Diehl, Sebastian and Rico, Enrique and Baranov, Mikhail A. and Zoller, Peter},
  title = {Topology by dissipation in atomic quantum wires},
  journal = {Nat. Phys.},
  volume = {7},
  pages = {971--977},
  year = {2011},
  doi = {10.1038/nphys2106},
  url = {https://www.nature.com/articles/nphys2106}
}

@article{Bardyn2013,
  author = {Bardyn, C.-E. and Baranov, M. A. and Kraus, C. V. and Rico, E. and {\.I}mamo{\u g}lu, A. and Zoller, P. and Diehl, S.},
  title = {Topology by dissipation},
  journal = {New J. Phys.},
  volume = {15},
  number = {8},
  pages = {085001},
  year = {2013},
  doi = {10.1088/1367-2630/15/8/085001},
  url = {https://arxiv.org/abs/1302.5135}
}

@article{Budich2015,
  author = {Budich, Jan Carl and Zoller, Peter and Diehl, Sebastian},
  title = {Dissipative preparation of {Chern} insulators},
  journal = {Phys. Rev. A},
  volume = {91},
  number = {4},
  pages = {042117},
  year = {2015},
  doi = {10.1103/PhysRevA.91.042117},
  url = {https://doi.org/10.1103/PhysRevA.91.042117}
}

@article{Poli2015,
  author = {Poli, Charles and Bellec, Matthieu and Kuhl, Ulrich and Mortessagne, Fabrice and Schomerus, Henning},
  title = {Selective enhancement of topologically induced interface states in a dielectric resonator chain},
  journal = {Nat. Commun.},
  volume = {6},
  pages = {6710},
  year = {2015},
  doi = {10.1038/ncomms7710},
  url = {https://www.nature.com/articles/ncomms7710}
}

@article{Weimann2017,
  author = {Weimann, S. and Kremer, M. and Plotnik, Y. and Lumer, Y. and Nolte, S. and Makris, K. G. and Segev, M. and Rechtsman, M. C. and Szameit, A.},
  title = {Topologically protected bound states in photonic parity--time-symmetric crystals},
  journal = {Nat. Mater.},
  volume = {16},
  number = {4},
  pages = {433--438},
  year = {2017},
  doi = {10.1038/nmat4811},
  url = {https://www.nature.com/articles/nmat4811}
}

@article{R-Tang2025,
  author = {Tang, Yugan and Wu, Jien and Lai, Pengtao and Hu, Yejian and Liu, Hui and Deng, Weiyin and Cheng, Hua and Liu, Zhengyou and Chen, Shuqi},
  title = {Loss induced delocalization of topological boundary modes},
  journal = {Nat. Commun.},
  volume = {16},
  pages = {10420},
  year = {2025},
  doi = {10.1038/s41467-025-65422-7},
  url = {https://www.nature.com/articles/s41467-025-65422-7}
}

@article{Sarkar2026,
  author = {Sarkar, Ronika and Hegde, Suraj S. and Narayan, Awadhesh and Meng, Tobias},
  title = {Chiral damping with persistent edge states from the interplay of topologies in open quantum systems},
  journal = {Commun. Phys.},
  volume = {9},
  pages = {109},
  year = {2026},
  doi = {10.1038/s42005-026-02573-z},
  url = {https://www.nature.com/articles/s42005-026-02573-z}
}

@article{Peng2026,
  author = {Peng, Yi and Yang, Chao and Hu, Haiping and Wang, Yucheng},
  title = {Dissipation-assisted preparation of topological edge states},
  journal = {Phys. Rev. B},
  volume = {114},
  number = {8},
  pages = {084203},
  year = {2026},
  doi = {10.1103/283r-6qnz},
  url = {https://journals.aps.org/prb/abstract/10.1103/283r-6qnz}
}

@article{R-Hatsugai,
  author = {Hatsugai, Yasuhiro},
  title = {{Chern} number and edge states in the integer quantum {Hall} effect},
  journal = {Phys. Rev. Lett.},
  volume = {71},
  pages = {3697--3700},
  year = {1993},
  doi = {10.1103/PhysRevLett.71.3697},
  url = {https://journals.aps.org/prl/abstract/10.1103/PhysRevLett.71.3697}
}

@article{Feshbach1962,
  author = {Feshbach, Herman},
  title = {A unified theory of nuclear reactions. {II}},
  journal = {Ann. Phys.},
  volume = {19},
  pages = {287--313},
  year = {1962},
  doi = {10.1016/0003-4916(62)90221-X},
  url = {https://www.sciencedirect.com/science/article/pii/000349166290221X}
}

@article{Rotter2009,
  author = {Rotter, Ingrid},
  title = {A non-{Hermitian} {Hamilton} operator and the physics of open quantum systems},
  journal = {J. Phys. A: Math. Theor.},
  volume = {42},
  number = {15},
  pages = {153001},
  year = {2009},
  doi = {10.1088/1751-8113/42/15/153001},
  url = {https://arxiv.org/abs/0711.2926}
}

@article{Dembowski2001,
  author = {Dembowski, C. and Gr{\"a}f, H.-D. and Harney, H. L. and Heine, A. and Heiss, W. D. and Rehfeld, H. and Richter, A.},
  title = {Experimental Observation of the Topological Structure of Exceptional Points},
  journal = {Phys. Rev. Lett.},
  volume = {86},
  number = {5},
  pages = {787--790},
  year = {2001},
  doi = {10.1103/PhysRevLett.86.787},
  url = {https://journals.aps.org/prl/abstract/10.1103/PhysRevLett.86.787}
}

@article{Yin2026Valley,
  author = {Yin, Shunda and Zhou, Qiuyan and Xi, Yuxiang and Deng, Weiyin and Chen, Wei and Lu, Jiuyang and Ke, Manzhu and Liu, Zhengyou},
  title = {Lossy phononic metamaterials for valley nonreciprocity},
  journal = {Nat. Commun.},
  volume = {17},
  pages = {3428},
  year = {2026},
  doi = {10.1038/s41467-026-70037-7},
  url = {https://doi.org/10.1038/s41467-026-70037-7}
}

@article{Jiang2024BoundaryHSTE,
  author = {Jiang, Tianshu and Zhang, Chenyu and Zhang, Ruo-Yang and Yu, Yingjuan and Guan, Zhenfu and Wei, Zeyong and Wang, Zhanshan and Cheng, Xinbin and Chan, C. T.},
  title = {Observation of non-{Hermitian} boundary induced hybrid skin-topological effect excited by synthetic complex frequencies},
  journal = {Nat. Commun.},
  volume = {15},
  pages = {10863},
  year = {2024},
  doi = {10.1038/s41467-024-55218-6},
  url = {https://doi.org/10.1038/s41467-024-55218-6}
}

@article{Xie2025ValleyLifetime,
  author = {Xie, Xinrong and Ma, Fei and Rui, W. B. and Dong, Zhaozhen and Du, Yulin and Xie, Wentao and Zhao, Y. X. and Chen, Hongsheng and Gao, Fei and Xue, Haoran},
  title = {Non-{Hermitian} {Dirac} cones with valley-dependent lifetimes},
  journal = {Nat. Commun.},
  volume = {16},
  pages = {1627},
  year = {2025},
  doi = {10.1038/s41467-025-56882-y},
  url = {https://doi.org/10.1038/s41467-025-56882-y}
}

@article{Fan2023Reconfigurable,
  author = {Fan, Haiyan and Gao, He and Liu, Tuo and An, Shuowei and Kong, Xianghong and Xu, Guoqiang and Zhu, Jie and Qiu, Cheng-Wei and Su, Zhongqing},
  title = {Reconfigurable topological modes in acoustic non-{Hermitian} crystals},
  journal = {Phys. Rev. B},
  volume = {107},
  pages = {L201108},
  year = {2023},
  doi = {10.1103/PhysRevB.107.L201108},
  url = {https://doi.org/10.1103/PhysRevB.107.L201108}
}

@article{Wang2022Morphing,
  author = {Wang, Wei and Wang, Xulong and Ma, Guancong},
  title = {Non-{Hermitian} morphing of topological modes},
  journal = {Nature},
  volume = {608},
  pages = {50--55},
  year = {2022},
  doi = {10.1038/s41586-022-04929-1},
  url = {https://doi.org/10.1038/s41586-022-04929-1}
}

@article{Slootman2024Monomode,
  author = {Slootman, E. and Cherifi, W. and Eek, L. and Arouca, R. and Bergholtz, E. J. and Bourennane, M. and Morais Smith, C.},
  title = {Breaking and resurgence of symmetry in the non-{Hermitian} {Su-Schrieffer-Heeger} model in photonic waveguides},
  journal = {Phys. Rev. Research},
  volume = {6},
  pages = {023140},
  year = {2024},
  doi = {10.1103/PhysRevResearch.6.023140},
  url = {https://doi.org/10.1103/PhysRevResearch.6.023140}
}

@article{Wu2024Helical,
  author = {Wu, Junpeng and Zheng, Riyi and Liang, Jialuo and Ke, Manzhu and Lu, Jiuyang and Deng, Weiyin and Huang, Xueqin and Liu, Zhengyou},
  title = {Spin-Dependent Localization of Helical Edge States in a Non-{Hermitian} Phononic Crystal},
  journal = {Phys. Rev. Lett.},
  volume = {133},
  pages = {126601},
  year = {2024},
  doi = {10.1103/PhysRevLett.133.126601},
  url = {https://doi.org/10.1103/PhysRevLett.133.126601}
}

@article{He2026CSE,
  title = {Non-Hermitian chiral skin effect of topological edge states},
  author = {Chu, Xiao-Meng and Wang, Xiao-Ran and Liang, Ying and He, Jing and Kou, Su-Peng},
  journal = {Phys. Rev. B},
  volume = {114},
  issue = {5},
  pages = {055102},
  numpages = {10},
  year = {2026},
  month = {Jul},
  publisher = {American Physical Society},
  doi = {10.1103/gdcf-ty8d},
  url = {https://link.aps.org/doi/10.1103/gdcf-ty8d}
}
\end{document}